\documentclass{iaus}
\usepackage{graphicx}
\usepackage{AMSsymb}

\title[Probing Orion Source I on 10-100 AU Scales] 
{A Documentary of High-Mass Star Formation: Probing the Dynamical
Evolution of Orion Source I on 10-100~AU Scales using SiO Masers}

\author[Matthews et al.]   
{L. D. Matthews$^1$,
 C. Goddi$^1$, L. J. Greenhill$^1$, C. J. Chandler$^2$, \break M. J. Reid$^1$,
\and E. M. L. Humphreys$^1$}

\affiliation{$^1$Harvard-Smithsonian Center for Astrophysics, 
Cambridge, MA, USA \\[\affilskip]
$^2$National Radio Astronomy Observatory, Socorro, NM, USA}

\pubyear{2007}
\volume{242}  
\pagerange{}
\date{?? and in revised form ??}
\jname{Proceedings of Astrophysical Masers and their Environments}
\editors{J. Chapman \& W. Baan, eds.}
\begin{document}

\maketitle

\begin{abstract}
A comprehensive picture of high-mass star formation 
has remained elusive, in part because examples of high-mass
young stellar objects (YSOs) tend to be relatively distant, deeply
embedded, and confused
with other emission sources. These factors have impeded 
dynamical investigations within tens of AU of high-mass YSOs---scales that
are critical for probing the interfaces where outflows from
accretion disks are launched and collimated. Using observations of 
SiO masers obtained with the Very Large Array (VLA) and the 
Very Long Baseline Array (VLBA), the KaLYPSO project 
is overcoming these limitations by 
mapping the structure and dynamical/temporal evolution of the 
material 10-1000 AU from the nearest high-mass YSO: Radio Source~I 
in the Orion BN/KL region. Our data include $\sim$40 epochs of 
VLBA observations over a several-year period, allowing us to track the proper 
motions of individual SiO maser spots and to monitor changes in the
physical conditions of the emitting material with time. Ultimately 
these data will provide 3-D maps of the outflow structure over approximately
30\% of the outflow crossing time. 
Here we summarize recent results from the KaLYPSO project, including
evidence that high-mass star formation occurs
via disk-mediated accretion. 

\keywords{masers, stars: pre-main sequence, techniques: high angular
resolution, circumstellar matter, stars: formation, stars: winds,
outflows, stars: individual (Source I)}
\end{abstract}

\firstsection 
\section{The KaLYPSO Project: Introduction and Motivation}

Increasing observational evidence supports a picture
in which high-mass stars ($M\gtrsim 10M_{\odot}$)
form through disk-mediated accretion, in a manner analogous
to a scaled-up version of low-mass star formation
(e.g., Zhang et al. 1998; Cesaroni 2002; Beuther et al. 2002; Patel
et al. 2005; Beltr\'an et al. 2006). However, the details of this process
remain poorly understood. For example, rapid formation timescales ($<<10^{6}$ years)
seem to necessitate extremely high mass accretion rates. Meanwhile, strong
radiation
pressure from a high-mass YSO is expected to impede
accretion, and the lack of a strong stellar
dynamo in high-mass protostars necessitates an alternative
mechanism for shedding large amounts of angular momentum. Although
a number of theoretical studies have attempted to address these issues
(e.g., McKee et al. 2002; Bonnell et al. 2005, Krumholtz et al. 2005),
observations  have been unable to
readily distinguish between various models. Open  questions
include: What are the sizes and structures of accretion disks around
high-mass YSOs?
What is the driving mechanism of their outflows?
What are the physical properties (density, temperature) of the
disk-outflow interface? What is the role of magnetic fields in 
high mass star formation?

One of the chief obstacles for constraining models of
high-mass star formation is that
examples of high-mass
YSOs tend to be relatively distant ($d\gtrsim$1~kpc), deeply embedded, and confused
with other emission sources. Moreover, since high-mass stars evolve
rapidly, by the time an unobstructed view of the young star emerges,
the disk
and outflow structures may have been destroyed.
Consequently, observations to date have been unable
to probe the 10-100~AU spatial scales over which outflows from
the accretion disks are
expected to be launched and collimated.
The KaLYPSO (Kleinmann-Low Young
Proto-Stellar Object) project (http://www.cfa.harvard.edu/kalypso/)
aims to overcome these limitations and provide the most detailed
picture yet of a high-mass star in  formation. 

\section{Our Target: Radio Source I in the Orion BN/KL Region}
The nearest region of high-mass star formation is the
Kleinmann-Low (KL) nebula in Orion, at a distance of $\sim$450~pc.
Within this nebula is located
the radio continuum-emitting ``Source~I'', a luminous, highly embedded
YSO with a disk-like morphology (Reid et al. 2007; Figure~1).

\section{Goals of the KaLYPSO Project}
The KaLYPSO team has launched a multi-faceted study of Source~I
with unprecedented angular and temporal resolution. Our
goals include:

\begin{itemize}

\item Chart the time-varying distribution of $\sim$1000 SiO maser
spots within 10-100~AU of Source~I

\item Measure proper motions of $\sim$100 individual SiO masers with
precision $<$1~km s$^{-1}$.

\item Produce geometric and dynamical models of the molecular gas
surrounding Source~I

\item Compare distributions of different maser transitions surrounding
Source~I to probe local density and temperature conditions

\item Constrain the physical processes (e.g., shocks,
collisions, fragmentation, magnetic fields) that contribute to the
process of forming a high-mass star

\item Produce a movie documenting the 3-D evolution of the outflows
surrounding a massive protostar over $\sim$30\% of the outflow crossing time

\end{itemize}

\section{The Observations of SiO Masers toward Source I}
Observations of masers using Very Long Baseline Interferometry
offer a powerful means to study
massive star formation. Masers are not affected by extinction, furnish
information on the physical conditions of the emitting gas, and can supply
kinematic information on the material surrounding YSOs
at extremely high angular resolutions.

Using the VLBA of the National Radio
Astronomy Observatory,
we have obtained $\sim$40 epochs of observations of the
SiO $v=1$ and $v$=2 masers toward Source~I at one-month intervals
over a multi-year period. The two transitions were
observed simultaneously with a spectral resolution of 0.2~km~s$^{-1}$. Our resulting
images have an angular resolution of $\sim$0.2~mas---corresponding to
$\sim$0.1~AU at the distance of Orion.

\section{Preliminary Results}
\subsection{The Large-Scale Picture}
Figure~1 shows
a large-scale ($\sim$1000~AU) bipolar structure surrounding Radio
Source~I, as previously observed in thermal and maser emission from the SiO ground
state ($v=0$) with $\sim0.5''$ resolution (Wright et al. 1995).
Spectroscopy  has confirmed similar velocity spreads in both
the north and south lobes, consistent with an outflow along the plane of the sky.
H$_{2}$O masers also lie along the apparent outflow.

\begin{figure}
\centering
\includegraphics[height=4in]{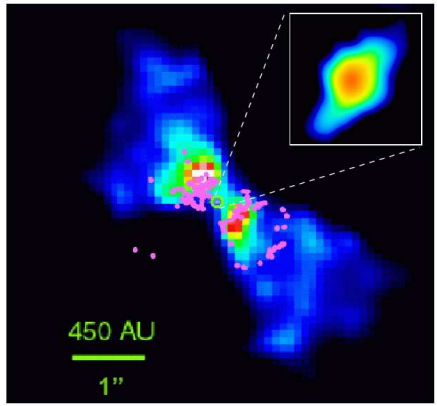}
\vspace{-0.7in}
  \caption[]{The large-scale ($\sim$1000~AU) bipolar structure surrounding Radio
Source~I, as observed in thermal and maser emission from the SiO ground
state ($v=0$) with $\sim0.5''$ resolution by Wright et al. 1995.
Locations of H$_{2}$O masers mapped with the VLA by Greenhill et al. 1998
are overplotted as pink spots. The inset shows
a $\lambda$7mm continuum map of Radio Source~I from Reid et al. 2007.
}\label{fig:bowtie}
\end{figure}

The Figure~1 inset shows 
a $\lambda$7mm continuum map of Radio Source~I from Reid et al. (2007).
The 7-mm emission appears to trace an edge-on, ionized disk with a
diameter of $\sim$100~AU, centered on the larger-scale outflow.
The SiO $v=1$ and $v=2$ observations described here are providing 
key new information linking the 
observations shown in Figure~1,
and further solidify the picture of Source~I as a high-mass star
in the process of formation via disk-mediated accretion and outflow.

\subsection{The Distribution and Kinematics of the SiO Masers around
Source~I}
Figure~2 shows a velocity field derived from one epoch of our
combined VLBA SiO
$v=1$ and $v=2$ observations toward Source~I. The maser emission lies
within an
``X''-shaped distribution centered on Source~I (whose position is
indicated by an asterisk symbol). The SiO $v=2$ emission tends to lie
closer to Source~I than the $v=1$ emission, although there is
considerable overlap. Emission is observed over
the velocity range
$-15\lesssim V_{\rm LSR}\lesssim30$~km~s$^{-1}$ (and is color-coded by Doppler
shift in the electronic version of this article). 
Emission to the north/west is predominantly
redshifted, while that to the east/south is predominantly
blue-shifted, suggesting rotational motion. Consistent with this,
material in the ``bridge'' between the south and west lobes
lies near the systemic velocity. A similar but fainter bridge between the north
and southeast lobes is also seen in some epochs. Proper motions of
individual maser features ($\sim$10~km~s$^{-1}$) 
are observed between consecutive epochs; their bulk directions are
indicated by arrows. A more detailed analysis of these proper motions
is forthcoming.

\begin{figure}
\centering
\includegraphics[height=3in]{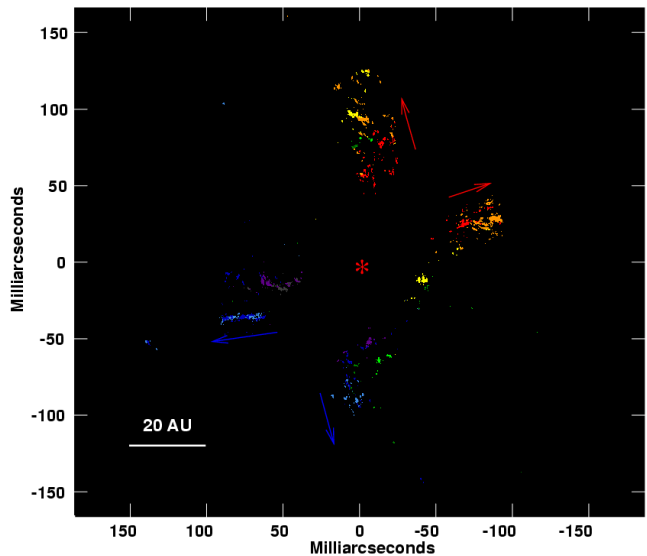}
  \caption[]{A velocity field derived from the
combined SiO
$v=1$ and $v=2$ emission toward Source~I during one observational
epoch. Emission is color-coded by Doppler shift (electronic version
only) and spans the velocity range $-15\lesssim V_{\rm
LSR}\lesssim30$~km~s$^{-1}$. 
An asterisk indicates the position of Source~I as determined
from radio continuum observations.
Directions of the bulk proper motions of individual maser spots
observed over the course of multiple observing epochs are indicated by
arrows.
}\label{fig:mom1}
\end{figure}

\section{An Emerging Model for Source~I}
The morphology and kinematics of the emission surrounding
Source~I provide strong evidence of both accretion and outflow.
Figure~3 shows a schematic model for
the source, based on  a synthesis of the observations
in the previous two figures (see the electronic version for a
full-color rendition).
At $r<150$~AU (dashed circle) material is being driven into
a wide-angle bipolar outflow. The edges of the ``funnel''
(representing the loci of the Doppler-shifted
SiO $v=1$ \& $v=2$ emission; see Figure~2)
may be infalling, rotating molecular material swept up by a wind/outflow
(Cunningham et al. 2005), or may demarcate the
edge-brightened portion of the wind from a
highly inclined ($i\sim80^{\circ}$),
flared, rotating accretion disk (traced at smaller radii
by 7-mm continuum
emission; see Figure~1). The canting of the individual SiO
emission arms (see Figure~2) and the existence of a ``bridge'' of
gas between the two sides of the funnel with a velocity
gradient across it (see Figure~2) now seem to exclude a model
of the masers as a biconical outflow along the northwest/southeast
direction (cf. Greenhill et al. 1998; Doeleman et al. 1999).

At $r>150$~AU (outside the dashed circle), cooler gas traced by
SiO $v=0$ maser+thermal emission (represented by the shaded areas; cf.
Figure~1) and H$_{2}$O
maser emission (depicted as white spots) delineate a more extended bipolar
outflow along a direction close to the plane of the sky.

\section{Summary}
Using the VLBA, our KaLYPSO team has obtained $\sim$40 epochs of monthly
observations of the SiO $v=1$ and $v=2$ masers toward Radio Source~I
in the Orion BN/KL region. Source~I is believed to be the nearest
example of a high-mass YSO. These observations are the first
to probe the kinematics of the molecular gas within 10-100~AU of such
an object.  Our analysis of these data, together with complementary
observations of the adjacent 7-mm radio continuum, H$_{2}$O masers, and SiO $v=0$
masers/thermal emission, provide compelling evidence that
disk-mediated accretion and outflow are fundamental to the process of
high-mass star formation. Further data reduction and analysis are
ongoing. Updates will be posted at 
http://www.cfa.harvard.edu/kalypso.

\begin{figure}
\centering
\includegraphics[height=3in]{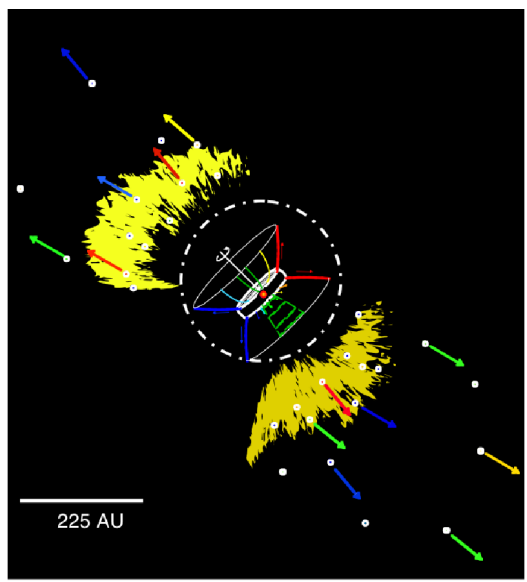}
  \caption[]{Schematic showing a working model for Source~I. Details
are described in the Text.
}\label{fig:cartoon}
\end{figure}

\begin{acknowledgments}
The Very Long Baseline Array (VLBA) of
the National Radio Astronomy
Observatory (NRAO) is a facility of the
National Science Foundation, operated under cooperative agreement by
Associated Universities, Inc. This work is support by grant 0507478 from
the National Science Foundation. 

\end{acknowledgments}

\end{document}